\documentclass[onecolumn,reviewcopy]{autart}
\usepackage{amssymb}
\usepackage{amsmath}
\usepackage[round]{natbib} %when use ieee cls file, this must be commented
\usepackage{color}
\usepackage{graphics}
\usepackage{graphicx}
\usepackage{multirow}
\usepackage{algorithm,algorithmic}
\usepackage{savesym}
\savesymbol{AND}
\usepackage{caption}
\usepackage{subcaption}
\graphicspath{ {./images/} }

\let\labelindent\relax
\usepackage{enumitem}
\usepackage{tikz}
\usetikzlibrary{fit,shapes.misc}
\usetikzlibrary{automata,positioning,arrows}
\usetikzlibrary{decorations.pathreplacing}

\def \kkk{\color{black}}

\def \L{\mathcal{L}}
\def \E{\mathcal{E}}
\def \V{\mathcal{V}}

\def \N{\mathcal{N}}

\def \D{\mathcal{D}}

\def \B{\mathcal{B}}

\def \Z{\mathbb{Z}}
\def \W{\mathcal{W}}

\def \a{\alpha}
\def \o{\omega}

\newtheorem{theorem}{Theorem}
\newtheorem{remark}{Remark}
\newtheorem{lemma}{Lemma}
\newtheorem{definition}{Definition}
\newtheorem{corollary}{Corollary}

\begin{document}
\maketitle
\thispagestyle{empty}
\pagestyle{empty}	
\begin{frontmatter}

\title{\Large \bf Consensus of Homogeneous Agents with General Linear Dynamics under Switching Communication Networks}

%\thanks[footnoteinfo]{Corresponding author}

\author[AddA]{Chong Jin Ong}\ead{mpeongcj@nus.edu.sg},
\author[AddA]{Ilayda Canyakmaz}\ead{ilayda.canyakmaz@u.nus.edu}

\address[AddA]{Department of Mechanical Engineering, National University of Singapore, 117576, Singapore}
%%\address[AddB]{Department of Mechanical Engineering, National University of Singapore, 117576, Singapore}

% \author{Chong Jin Ong\corref{cor1}\fnref{label1}}
% \ead{mpeongcj@nus.edu.sg}
% %\ead[url]{home page}
% \fntext[label1]{}
% \cortext[cor1]{}
% \affiliation{organization={Department of Mechanical Engineering, National University of Singapore},
%             %addressline={},
%            % city={},
%             postcode={117576},
%             %state={},
%             country={Singapore}}
% %\fntext[label3]{}
% \author{Ilayda Canyakmaz\fnref{label1}}
% \ead{ilayda.canyakmaz@u.nus.edu}
% %\ead[url]{home page}
%% \fntext[label2]{}
%% \affiliation{organization={Department of Mechanical Engineering, National University of Singapore},
%             %addressline={},
%            % city={},
%             % postcode={117576},
%             %state={},
%             % country={Singapore}}
% %\fntext[label3]{}
%\author{Chong Jin Ong\corref{cor1}\fnref{label1}}
% \ead{mpeongcj@nus.edu.sg}
%\author{Ilayda Canyakmaz\fnref{label1}}
% \ead{ilayda.canyakmaz@u.nus.edu;Ilayda_Canyakmaz@sutd.edu.sg}
%
%\affiliation[label1]{organization={Department of Mechanical Engineering, National University of Singapore},%Department and Organization
%            postcode={117576},
%            country={Singapore}}

\begin{abstract}
This work addresses the synchronization/consensus problem of identical multi-agent system (MAS) where the agents' dynamics are linear and the communication network is arbitrarily switching among connected topologies.  The approach uses a gain matrix of a special structure in a dynamic compensator for each agent. Under reasonable conditions, the approach ensures that consensus is reached when the gain is sufficiently large. This result holds for general linear systems including the case where agents have repeated unstable eigenvalues.  The proposed controller structure can be seen as a special case of the existing MAS controller structures but offers consensus conditions that are simpler than existing results, especially for the case when the network is switching among connected graphs.  The works shows its application to three communication settings: fixed graph, switching among undirected and connected graphs and switching among directed and connected graphs.  An example is provided to illustrate the results.
\end{abstract}

\begin{keyword}
Consensus, Multi-agent system, Network System.
\end{keyword}

\end{frontmatter}

\section{Introduction}

The study of multi-agent system (MAS) has been an active area of research and one well-studied topic in MAS is the consensus or synchronization (hereafter referred to collectively as consensus) problem among agents connected via a communication network. Early studies of this problem deal with agents that are scalar systems \citep{ART:RB05,ART:M05,INP:M04}, double integrator systems \citep{ART:R08} and recent focus is on agents that have general linear dynamics \citep{ART:SS09,ART:T09,ART:HWJ13,ART:VZ17}. Many of these works deal with the case where the communication network is fixed and connected \citep{ART:T09,ART:ZLD11,ART:HWJ13}.  When the network is switching, techniques of switching systems, like dwell-time requirement, or appropriate Lyapunov functions are then used to ensure consensus \citep{ART:LM99,ART:LA09a}. For example, \cite{ART:VZ17} shows such a result when the network switches among connected and undirected graphs for agents having a single input.

This work proposes an approach that does not follow such a procedure.  More precisely, the consensus problem under switching networks is based on the closed-form expression of the time solutions of the MAS without explicit dwell-time/Lyapunov requirement.  It ensures consensus is reached when the network switches among connected graphs (for both undirected and directed) in a leaderless network for general linear system. The controller uses a gain matrix, obtained from a diagonal structure, multiplied into the standard diffusive term. The proposed approach has several useful features. Firstly, it uses a dynamic compensator and works for agents with multiple inputs and when not all states are measurable. Secondly, the gain matrix can be obtained constructively and is guaranteed to exist under mild conditions for a fixed graph without the use of Riccatti/Lyapunov equations. Consensus is achieved when a condition involving the second largest eigenvalue of the Laplacian matrix is satisified. Thirdly, the proposed approach yields a straight forward condition under the \textit{same} framework for achieving consensus when the MAS system switches arbitrarily among connected and undirected graphs. In the case of arbitrary switching among directed and connected graphs, the conditions show that consensus can be achieved when the gain is sufficiently large so long as the network does not switch instantaneously. The proposed structure also offers interesting insights to achieving consensus using just the weights of the Laplacian matrix alone, see for example comments in Remark \ref{rem:gamma1}.

To the best of our knowledge, this special structure has not been attempted in the MAS literature. A shortened version of this work \citep{INP:OC22} has been accepted under the stronger assumption that $A$ is diagonalizable.  This work first shows how this assumption is relaxed, followed by results for achieving consensus under three different communication settings: the network is fixed, the network is switching among connected, undirected graphs and the network is switching among connected, directed graphs.

The rest of this paper is organized as follows. This section ends with a description of the notations used. Section \ref{sec:pre} reviews standard definitions, graph and matrix properties. Section \ref{sec:special} discusses the solutions of some special classes of agents. The intention is to simplify the subsequent results by avoiding the consideration of non-diagonalizable $A$.  Section \ref{sec:Proposed} shows the structure of the proposed controller. Section \ref{sec:fixedgraph} shows the necessary and sufficient conditions for consensus for the case of a fixed network MAS. Section \ref{sec:undirected} considers the case where the network switches among undirected and connected graphs. While section \ref{sec:directed} is for directed and connected graphs. An example that illustrates the stated results is given in Section \ref{sec:num}. The work concludes in Section \ref{sec:con}.

The notations used in this paper are standard. Non-negative and positive integer sets are $\mathbb{Z}^+_0$ and $\mathbb{Z}^+$ respectively. Selected ranges of the integer set are $\mathbb{Z}^N=\{1,\cdots, N\}$ and $\mathbb{Z}_{\ell}^k=\{\ell, \ell+1, \cdots, k\}$ with $k > \ell$. Similarly,  the sets of real numbers, $n$-dimensional real vectors and $n$ by $m$ real matrices are $\mathbb{R}, \mathbb{R}^n, \mathbb{R}^{n \times m}$ respectively. For a square matrix $Q$, $Q \succ (\succeq) 0$ means $Q$ is positive definite (semi-definite), $Q'$ is its transpose and $spec(Q)$ is its set of eigenvalues. The $2$-norm of $x \in \mathbb{R}^n$ is $\|x\|$, $I_n$ is the $n\times n$ identity matrix, $1_m$ ($\hat{1}_m$) is the $m$-vector of all 1 ($\frac{1}{\sqrt{m}}$),
$\mathbb{C}^- (\bar{\mathbb{C}}^-)$ and $\mathbb{C}^+ (\bar{\mathbb{C}}^+)$ are the open (closed) left half and the open (closed) right half of the complex plane respectively.  Given $A \in \mathbb{R}^{n \times m}, B \in \mathbb{R}^{q \times r}$,
$A \otimes B \in \mathbb{R}^{nq \times mr}$ is the Kronecker product of $A$ and $B$.
Given $x_i \in \mathbb{R}^n, i \in \mathbb{Z}^N$, $x^j_i \in \mathbb{R}$ is the $j^{th}$ element of $x_i$, the bold-faced symbols $\pmb{x}^q=[x^q_1 \: x^q_2 \: \cdots \: x^q_M]' \in \mathbb{R}^{M}, q \in \mathbb{Z}^n$ and $\hat{\pmb{x}}=[(\pmb{x}^1)' \cdots (\pmb{x}^n)']'=[ (x_1^1 x_2^1 \cdots x_M^{1}), (x_1^2 x_2^2 \cdots x_M^2),\cdots (x_1^n x_2^n \cdots x_M^n)]' \in \mathbb{R}^{Mn}$. Block diagonal matrix with $m$ diagonal blocks is denoted as $diag_m\{d_1,\cdots,d_m\}$ with $d_i \in \mathbb{R}^{n_i \times n_i}$ including the case where $n_i=1$. $\N^r \in \mathbb{R}^{r \times r}$ is the canonical nilpotent matrix having zero elements everywhere except the value of $1$ at the $(i,i+1)$ elements, $i \in \mathbb{Z}^{r-1}$.
%$$\N^r=\begin{bmatrix}
%    0&1 &0& \cdots &0 \\
%    0&0& 1 & \cdots &0\\
%    \vdots&\vdots & \vdots & \ddots& \vdots \\
%%    0&0 & 0 & \cdots & 1\\
%    0&0&0 & 0 & 0
%  \end{bmatrix} \textrm{ with its matrix exponential } e^{\N^r t} =\begin{bmatrix}
%    1 & t & \frac{t^2}{2!} & \cdots &\frac{t^{r-1}}{(r-1)!} \\
%    0&1& t & \cdots & \\
%    \vdots&\vdots & \vdots & \ddots& \vdots \\
%%    0&0 & 0 & \cdots & 1\\
%    0&0&0 & 0 & 1
%  \end{bmatrix}$$
Additional notations are introduced when required.

%%%%%%%%%%%%%%%%%%%%%%%%%%%%%%%%%%%%%%%%%%%%%%%%%%%%%%%%%%
\section{Prelimenaries} \label{sec:pre}
As in standard MAS, the communication among agents is given by the graph $G(t) = (\V,\E(t))$ with vertex set  $\V = \{1,2,\cdots,m\}$  and edge set $\E(t) \subseteq \V \times \V$. The notation $(i,j) \in \E(t)$ means that $j$ is a in-neighbor of $i$ at time $t$. This network is described by a $m \times m$ Laplacian matrix $\L(G(t))$ with elements $\L_{ij}(G(t))=-\a_{ij}(t)$ if $(i,j)\in \E(t)$ and $i \neq j$, $\L_{ii}(G(t))=\sum_{j=1}^m \a_{ij}(t)$ for all $i \in \Z^m$ and $\L_{ij}(G(t))=0$ otherwise.
Note that $\a_{ij}(t)$ is nonnegative for all $i,j \in \V \times \V$ and, if nonzero, is greater than $\bar{\a}$, for some $\bar{\a}>0$ \citep{INP:M04}.
A few common definitions of connectivity of $G$ are reviewed and stated next, together with a summary of well-known properties of $\L$.
%For notational convenience, $\L(G)$ is referred to as a connected matrix when $G$ is a connected graph.
\begin{definition}
A graph, $G$, is connected if there exists a base node such that all other nodes can be reached from the base node via the edges of the graph.
\end{definition}
%A well-known fact of a connected graph is that $G(t)$ contains a spanning tree.
\begin{definition}
A directed graph $G$ is strongly connected if there exists a path, following the edges of the graph, that connects every pair of nodes.
\end{definition}
%\rrr Not needed. \begin{definition}
%A graph $G(t)$ is uniformly connected if there exists $\tau>0$ such that the union of the graph $G(t)$ from $t$ to $t+\tau$, defined by $G(\V, \cup_{k=t}^{k=t+\tau}\E(k))$, is connected for all $t$.
%\rrr Is this for undirected graph only? Need strongly connected for directed $G$? \kkk
%\end{definition}
\kkk
Several well-known properties of $\L(G)$ include
\begin{enumerate}[label=(D\arabic*),itemindent=0.5em]
  \item The sum of each row of $\L$ is 0.
  \item Smallest eigenvalue of $\L$ is 0 with right eigenvector of $1_m$.
  \item When $G$ is connected (or strongly connected), the smallest eigenvalue of $\L(G)$ is simple.
  \item All eigenvalues of $\L$ lie in the closed right half of the complex plane. In the case where $G$ is undirected, $\L(G)$ is symmetric and hence has all real eigenvalues.
\end{enumerate}
Suppose $A \in \mathbb{R}^{r \times n}$, $B \in \mathbb{R}^{p \times q}$ and $C,D$ are of appropriate dimensions. Then \citep{BOO:HJ91}
%two useful properties of the kronecker product are
\begin{enumerate}[label=(P\arabic*),itemindent=0.5em]
%  \item[(P1)] $A \otimes I_m + B \otimes I_m = (A+B) \otimes I_m $
%  \item[(P2)] $AB \otimes I_m = (A \otimes I_m)(B \otimes I_m)$
  \item $A \otimes B = (A \otimes I_p)(I_n \otimes B)=(I_r \otimes B)(A \otimes I_q)$
  \item $ AB \otimes CD = (A \otimes C)(B \otimes D)$
\kkk
\end{enumerate}
If $A, B \in \mathbb{R}^{n \times n}$, then
%Additional properties can be derived from the above. Specifically, given $A \otimes I_m$ where $A \in \mathbb{R}^{n \times n}$, then using (P2) repeatedly and the series expansion of the matrix exponential,
%\begin{align*}
%e^{(A \otimes I_m)t}&= I_n \otimes I_m + A \otimes I_m t + (A \otimes I_m)^2\frac{t^2}{2!} + \cdots\\
%&= I_n \otimes I_m + A \otimes I_m t + A^2\frac{t^2}{2!} \otimes I_m + \cdots = e^{At} \otimes I_m.
%\end{align*}
%A similar result can be obtained for $e^{(I_m\otimes A)t}$.  Hence,
\begin{enumerate}[label=(P\arabic*),itemindent=0.5em]\setcounter{enumi}{2}
  \item $e^{(A+B)t}=e^{At}e^{Bt}$ if and only if $A B=B A$.
  \item $e^{(A \otimes I_m)t}= e^{At} \otimes I_m$.
  \item $e^{(I_m \otimes A)t}= I_m \otimes e^{At}$.
  \item If $(\lambda, v)$ is the eigenvalue and eigenvector of $A$, then $(e^{\lambda t},v)$ is the corresponding eigenvalue and eigenvector of $e^{At}$.
\end{enumerate}

\section{Special Classes of Agents}\label{sec:special}
Before stating the most general form of the proposed control law, some special cases are first considered. For this purpose, the notations for switching communication network is first established. Let

\textbf{(A1)} The communication graph $G(t)$ switches as instants $t_0, t_1, \cdots $ and that $t_{k+1}-t_k \geq \underline{h}$ for some dwell time $\underline{h}>0$.

Three simple results are first shown for a special case of the proposed approach, to be followed by results for the case of a general $A$.
%These two results are for the cases when $A=I_n$ in (\ref{eqn:dotxi}) and $\dot{x}_i(t)=J_n x_i(t)$ where $J_n$ is the Jordan block of order $n$.
The first is for the case the agent is such that
\begin{align*}
   \dot{x}_i(t)=& \gamma \sum_{j=1}^m \a_{ij}(t) ({x}_j(t) - {x}_i(t)), i \in \mathbb{Z}^m %\label{eqn:dotxi2}
\end{align*}
where $x_i \in \mathbb{R}^{n}$, $\a_{ij} \in \mathbb{R}$ is the weighing coefficients and $\gamma > 0$ is some constant.  This, when expressed as $\hat{\pmb{x}}=[(\pmb{x}^1)' \cdots (\pmb{x}^n)']' \in \mathbb{R}^{nm}$ becomes
\begin{align}\label{eqn:dothatpmbx}
   \dot{\hat{\pmb{x}}}(t)=  I_n \otimes (-\gamma \L(t))\hat{\pmb{x}}(t)
\end{align}
with $\L(t):=\L(G(t))$. Under (A1), it is clear that $G(t)=G(t_k)$ for $t \in [t_k, t_{k+1})$. Hence,
\begin{align}
\L(t):=\L_k \textrm{ for } t \in [t_k, t_{k+1}) \textrm{ and } h_k:=t_{k+1}-t_k. \label{eqn:Lt}
\end{align}
Then the solution, using (P5), is
\begin{align}
%\hat{\pmb{x}}(t_{k+1})&= e^{\gamma h_k} e^{ (I_n \otimes (-\L_k))h_k}\hat{\pmb{x}}(t_k)= e^{\gamma h_k} (I_n \otimes e^{-\L_k h_k})\hat{\pmb{x}}(t_k)\nonumber \\%\label{eqn:hatpmbx}
\hat{\pmb{x}}(t_{k+1}) &= (I_n \otimes e^{-\gamma \L_k h_k})\hat{\pmb{x}}(t_k). \label{eqn:hatpmbx}
\end{align}
The second result is that for $\dot{x}_i(t)=J_n x_i(t), i \in \mathbb{Z}^m$ where $J_n$ is an $n-$order Jordan block with eigenvalue $\lambda$ in the sense that $J_n=\lambda I_n + \N^n$ where $\N^n$ is the $n$-order canonical nilpotent matrix. Similarly, the MAS can be expressed as
\begin{align}\label{eqn:dothatpmbx3}
  \dot{\hat{\pmb{x}}}(t)=(J_n \otimes I_m)\hat{\pmb{x}}(t)
\end{align}
which, since $\lambda I_n$ and $\N^n$ commute, has a solution under (P3) and (P4) as
\begin{align}
 \hat{\pmb{x}}(t)&=((e^{J_n t}) \otimes I_m)\hat{\pmb{x}}(0)
 %= ((e^{(\lambda I_n +\N^n) t}) \otimes I_M)\hat{\pmb{x}}(0)= ((e^{\lambda t}e^{\N^n t}) \otimes I_M)\hat{\pmb{x}}(0)\nonumber\\
    = e^{\lambda t}e^{\N^n t} \otimes I_m \hat{\pmb{x}}(0). \label{eqn:hatpmbx2}
\end{align}
The third result is one that combines the above two, stated as a lemma for easy reference.
\begin{lemma}\label{lem:dotxi}
Let $J_n= \lambda I_n + \N^n$. The MAS
\begin{align}\label{eqn:dotxit}
\dot{x}_i(t)=J_n x_i(t)+ \gamma I_n \sum_{j=1}^m \a_{ij}(t) ({x}_j(t) - {x}_i(t)), \: i \in \mathbb{Z}^m
\end{align}
reaches consensus if the MAS
\begin{align}
\dot{x}_i(t)=\lambda I_n x_i(t)+ \gamma I_n  \sum_{j=1}^m \a_{ij}(t) ({x}_j(t) - {x}_i(t)), \: i \in \mathbb{Z}^m\label{eqn:dotxi3}
\end{align}
reaches consensus exponentially for all $x_i(0), i\in \mathbb{Z}^m$.
\end{lemma}

The above result shows clearly that the value of $\gamma$ that achieves consensus for (\ref{eqn:dotxi3}) also does so for \ref{eqn:dotxit}. In what follows, this fact is used to for the design of $\gamma$ for systems with $A$ being non-diagonalizable.

\section{The proposed controller and the closed-loop structure}\label{sec:Proposed}
This section considers the general case of unstable $A$.
%The agent dynamics, together with the proposed controller has a very similar structure to those used in the literature \cite{ART:SS09} except for the use of a weighted matrix $\Phi$.
For avoidance of notational conflict, let
\begin{align}\label{eqn:dotziopenloop}
\dot{w}_i(t)= A w_i(t)+B u_i(t)
\end{align} \kkk
be the open-loop dynamics of the $i$-th agent with $\o_i \in \mathbb{R}^{n}$ being its states, $B \in \mathbb{R}^{n \times n_B}, n_B < n$ and assume that\\
\textbf{(A2)} $(A, B)$ is stabilizable.\\
\textbf{(A3)} All states are measurable.\\
The controller is a dynamical compensator of the form
\begin{align}
\dot{\eta}_i&=(A+B K)\eta_i+ \Phi \sum_j \a_{ij}(t)(\eta_j-\eta_i+w_i-w_j)\label{eqn:doteta1}\\
u_i&=K\eta_i\label{eqn:doteta2}
\end{align} \kkk
where $\Phi \in \mathbb{R}^{n \times n}$ is the controller gain, $\eta_i \in \mathbb{R}^{n}$ is the state of the compensator and
$K \in \mathbb{R}^{n_B \times n}$ is such that $(A+BK)$ is Hurwitz.
Let $z_i:=w_i-\eta_i$ and it follows from the combined system of (\ref{eqn:dotziopenloop})-(\ref{eqn:doteta2}) that
\begin{align}
%s_i(t)&:=x_i(t)-\eta_i(t)\\
\dot{z}_i&=\dot{w}_i-\dot{\eta}_i\nonumber\\&=Aw_i+B u_i-(A+BK)\eta_i-\Phi \sum_j \a_{ij}(t)(z_i - z_j)\nonumber \\
%&=A(z_i-\eta_i)+ \Phi \sum_j \a_{ij}(t)(s_j-s_i)
&=Az_i+ \Phi \sum_j \a_{ij}(t) (z_j-z_i) \label{eqn:dotzi}
\end{align} \kkk
If $z_i$ reaches consensus in (\ref{eqn:dotzi}), $\eta_i$ approaches $0$ from (\ref{eqn:doteta1}). Then $w_i$ reaches consensus since $w_i -w_j = z_i - z_j+(\eta_i -\eta_j)$ \kkk. Hence, the important result is to achieve consensus for the MAS having a structure of the form given by (\ref{eqn:dotzi}) where
$\Phi$ is an appropriate matrix.
%Clearly,  the above expression has the same structure as (\ref{eqn:dotzi}).
%It also becomes the standard system of (\ref{eqn:dotxi}) when $\Phi=I_n$.

%The consensus for $s_i$ is achieved if (\ref{eqn:dotxiS}) reaches consensus with the appropriate $\Gamma$.
%In view of this, the focus is on consensus conditions for the system of
%\begin{align}\label{eqn:dotzi}
%  \dot{z}_i= A z_i + \Phi \sum \a_{ij}(t)(z_j-z_i),  i \in \mathbb{Z}^m
%\end{align}
%where $\Phi$ is an appropriate matrix. Clearly,  the above expression has the same structure as (\ref{eqn:dotzi}). It also becomes the standard system of (\ref{eqn:dotxi}) when $\Phi=I_n$.

The most general result is to express $A$ via appropriate transformation into its Jordan form. However, due to the result of Lemma \ref{lem:dotxi}, this is not necessary. Instead, the following assumption is made:\\
\textbf{(A4)} $A$ is diagonalizable and has $r$ eigenvalues in $\mathbb{C}^+$.\\
Under this assumption, the expression of (\ref{eqn:dotzi}) can be transformed via $z_i=Qx_i, i \in \mathbb{Z}^m$ where $Q \in \mathbb{R}^{n\times n}$ contains column-wise eigenvectors of $A$ as
\begin{align}
\dot{x}_i(t)&=S x_i(t) + \Gamma \sum_{j=1}^m \a_{ij}(t) (x_j(t) - x_i(t)), \: i \in \mathbb{Z}^m \label{eqn:dotxiS}
\end{align}
where $\Gamma=Q^{-1}\Phi Q$ and
\begin{subequations} \label{eqn:S}
\begin{align}
S=Q^{-1}AQ&=diag_n\{\lambda_A^1, \cdots, \lambda_A^r \cdots, \lambda_A^n\}
\end{align}
\begin{align}
 \textrm{with } Re(\lambda_A^1) \geq \cdots \geq Re(\lambda_A^n)
\end{align}
\end{subequations}
\begin{align}
\{\lambda_A^1, \cdots, \lambda_A^r\} &\in \mathbb{C}^+ \textrm{ and } \{\lambda_A^{r+1}, \cdots, \lambda_A^n\} \in \bar{\mathbb{C}}^-. \label{eqn:specAinC}
\end{align}
Hereafter, all MASs are assumed to have the form of (\ref{eqn:dotxiS}) for simplicity in presentation.
Clearly, if (\ref{eqn:dotxiS}) reaches consensus, so does system (\ref{eqn:dotzi}).
While $\Gamma$ can be a general matrix, the proposed approach is to let
\begin{align}
\Gamma=diag_n\{ \gamma^1, \cdots, \gamma^n\} \label{eqn:Gamma}
\end{align}
for all agents. Then, since both $S$ and $\Gamma$ are diagonal, the MAS can be written as
\begin{align}\label{eqn:dothatpmbxt}
&\dot{\hat{\pmb{x}}}(t)= (S \otimes I_m - \Gamma \otimes \L(t)) \hat{\pmb{x}}(t) \nonumber\\
&= diag_n \{ \lambda_A^1 I_m - \gamma^1 \L(t),
 \cdots, \lambda_A^n I_m - \gamma^n \L(t)  \} \hat{\pmb{x}}(t)
\end{align}
The above is a block-diagonal system consisting of $n$ blocks with the $i$ block being
\begin{align}\label{eqn:dotpmbx}
\dot{\pmb{x}}^i(t) = (\lambda_A^i I_m - \gamma^i \L(t)) \pmb{x}^i(t), \quad i \in \mathbb{Z}^n
\end{align}
The next few sections discuss the results for this system under different switching conditions.

\begin{remark}\label{rem:zxconversion1}
The system of (\ref{eqn:dotxiS}) can also be obtained from (\ref{eqn:doteta1}) using $u_i= B' (BB')^{-1} \Phi \sum_{j=1}^m \a_{ij}(t) (w_j(t) - w_i(t))$ in the case where $B$ has full row rank.
\end{remark}

%\begin{remark}\label{rem:zxconversion2}
%In the typical case where $B$ is less than full row rank, our approach follows the suggestion of \cite{ART:SS09} in which a dynamical compensator is used having the form of
%\begin{align}
%\dot{\eta}_i&=(A+BK)\eta_i+ \Phi \sum \a_{ij}(t)(\eta_j-\eta_i+z_i-z_j)\label{eqn:doteta1}\\
%u_i&=K\eta_i\label{eqn:doteta2}
%\end{align}
%for system (\ref{eqn:dotzi}). By letting $s_i:=z_i-\eta_i$, it follows that
%\begin{align*}
%%s_i(t)&:=x_i(t)-\eta_i(t)\\
%\dot{s}_i&=\dot{z}_i-\dot{\eta}_i=Az_i+Bu_i-(A+BK)\eta_i-\Phi \sum \a_{ij}(t)(\eta_j-\eta_i+z_i-z_j)\\
%&=A(z_i-\eta_i)+ \Phi \sum \a_{ij}(t)(z_j-\eta_j-(z_i-\eta_i)) \\
%&=As_i+ \Phi \sum \a_{ij}(t) (s_j-s_i)
%\end{align*}
%which has the same structure as system (\ref{eqn:dotzi}). If $s_i$ reaches consensus, $\eta$ approaches $0$ from (\ref{eqn:doteta1}) which implies that $z_i$ reaches consensus.  The consensus for $s_i$ is achieved if (\ref{eqn:dotxiS}) reaches consensus with the appropriate $\Gamma$.
%\end{remark}
\begin{remark}\label{rem:outputfeedback}
In the case where some of the states are not measurable, an observer-based dynamic compensator can be used. The design is quite standard and, hence, only the essential steps are presented. Obviously, the additional assumption that $(A,C)$ is detectable is needed besides (A2).  The dynamic compensator in this case is given by
\begin{align*}
\dot{\tilde{w}}_i&=A\tilde{w}_i+Bu_i+H(\tilde{y}_i-y_i), \quad \tilde{y}_i =C\tilde{w}_i\\
\dot{\eta}_i&=(A+BK)\eta_i+ \Phi \sum \a_{ij}(t)(\eta_j-\eta_i+\tilde{w}_i-\tilde{w}_j)\nonumber\\ u_i&=K\eta_i
\end{align*}
Letting $z_i:=\tilde{w}_i-\eta_i$ and $e_i:=w_i-\tilde{w}_i$, it is easy to show that
\begin{align*}
% \dot{s}_i &=\dot{\tilde{z}}_i-\dot{\eta}_i=A\tilde{z}_i+BK\eta_i+HC(\tilde{z}_i-z_i)-(A+BK)\eta_i-\Phi \sum \a_{ij}(t)(\eta_j-\eta_i+\tilde{z}_i-\tilde{z}_j)\\
%  &=A(\tilde{z}_i-\eta_i)+\Phi \sum \a_{ij}(t)(\tilde{z}_j-\eta_j-(\tilde{z}_i-\eta_i))+HC(\tilde{z}_i-z_i)\\
\dot{z}_i &=Az_i+\Phi \sum \a_{ij}(t)(z_j-z_i)-HCe_i\\
%\dot{e}_i&=\dot{z}_i-\dot{\tilde{z}}_i =Az_i+Bu_i-A\tilde{z}_i-Bu_i-HC(\tilde{z}_i-z_i)\\
%&=A(z_i-\tilde{z}_i)+HC(z_i-\tilde{z}_i) =(A+HC)e_i
\dot{e}_i& =(A+HC)e_i
\end{align*}
Clearly, when $H$ is chosen such that $(A+HC)$ is Hurwitz, $e_i$ goes to zero and the last term of the $\dot{z}_i$ equation above goes to zero. When that happens, the expression of $z_i$ becomes that of (\ref{eqn:dotzi}) and $z_i$ reaches consensus if (\ref{eqn:dotxiS}) does.
\end{remark}

\begin{remark}\label{rem:nondiagonal}
In the event that $A$ is not diagonalizable, the choice of $Q$ is such that $Q^{-1}AQ=diag_p\{ \pmb{J^1}, \cdots  \pmb{J^p}\}$ where $\pmb{J^i}=diag_i\{J^i_1, \cdots J^i_{n_i}\}$ with each $J^i_p= \lambda_i I_{n_p}+ \N^{n_p}$ is a $n_p$-order Jordan matrix. The choice of $\Gamma$ of (\ref{eqn:Gamma}) can then be designed using the procedure above but based on $S=diag_p\{\lambda_1 I^1, \cdots, \lambda_p I^p\}$ where $I^i$ is the identity matrix of dimension equal to that of $\pmb{J^i}$. Using this choice of $\Gamma$ results in consensus of the original system following the result of lemma \ref{lem:dotxi}.
\end{remark}
%The above discussion shows that achieving consensus for (\ref{eqn:dotxiS}) is the key result. Clearly, when $x_i$ reaches consensus for (\ref{eqn:dotxiS}), $z_i=Qx_i$ implies that (\ref{eqn:dotzi}) reaches consensus for $z_i$. Hence, the discussions hereafter assume that the MAS system has the structure of (\ref{eqn:dotxiS}), unless otherwise stated.

\section{Consensus under fixed graph} \label{sec:fixedgraph}
This section considers the case when the communication network is fixed.
\begin{theorem}\label{thm:fixed}
Suppose (A4) holds and $G$ is a fixed, connected and directed graph. Let $\gamma^{r+1}=\cdots=\gamma^n = 1$. The MAS system of (\ref{eqn:dotxiS}) with conditions (\ref{eqn:S}),  (\ref{eqn:specAinC}) and (\ref{eqn:Gamma}) reaches consensus if and only if $\gamma^i > \frac{Re(\lambda_A^i)}{Re(\lambda^2_\L)}$ for $i \in \mathbb{Z}^r$ where $\lambda^2_{\L}$ is the second smallest eigenvalue of $\L$.
\end{theorem}
\textbf{Proof:} $(\Rightarrow)$ Since $S$ and $\Gamma$ are diagonal, system (\ref{eqn:dotxiS}) can be expressed as (\ref{eqn:dotpmbx}).
The solution of (\ref{eqn:dotpmbx}), following the fact that $\lambda_A^i I_m$ and $\gamma^i \L$ commute, is
%the solution of (\ref{eqn:dothatpmbx5}) rewritten as a diagonal matrix of $n$ blocks with the $i^{th}$ block being of the form
%$$ \pmb{x}^i(t)=e^{(\lambda_A^iI_m-\gamma^i \L) t} \pmb{x}^i(0)=e^{\lambda_A^i t} e^{-\gamma^i \L t} \pmb{x}^i(0), \: i \in \mathbb{Z}^n$$
\begin{align}
\pmb{x}^i(t)&=e^{(\lambda_A^iI_m-\gamma^i \L) t} \pmb{x}^i(0)=e^{\lambda_A^i t} e^{-\gamma^i \L t} \pmb{x}^i(0)\nonumber\\&= e^{\lambda_A^i t} (\sum_{j=1}^m \varphi_j \xi_j' e^{-\gamma^i  \lambda_\L^j t}) \pmb{x}^i(0) \nonumber \\% \: i \in \mathbb{Z}^n
%&= \sum_{j=1}^m  e^{(\lambda_A^1-\gamma^1 \lambda_\L^j)t} \varphi_j \xi_j'\pmb{x}^1(0)
&= e^{\lambda_A^i t} 1_m \xi_1' \pmb{x}^i(0) + \sum_{j=2}^m   e^{(\lambda_A^i -\gamma^i \lambda_\L^j)t} \varphi_j \xi_j'\pmb{x}^i(0) \label{eqn:pmbxi}
%&=(e^{(\lambda_s^1-\lambda_\L^1)t} \varphi_1 \xi_1'+e^{(\lambda_s^1-\lambda_\L^2)t} \varphi_2 \xi_2'+\cdots +e^{(\lambda_s^1-\lambda_\L^m)t} \varphi_m \xi_m')\pmb{x}^1(0)
\end{align}
where the eigen-decomposition of $e^{-\gamma^i \L t}$ is used and $\lambda_\L^j$ is the $j^{th}$ eigenvalue of $\L$ with right and left eigenvectors $\varphi_j$ and $\xi_j$ respectively. The last equality on the right above follows because $\L$ is a connected graph which by (D3) means that $\varphi_1 = 1_m$, $\lambda_\L^1=0$ and $Re(\lambda_\L^j) > 0$ for $j\in \mathbb{Z}_2^m$. When $\gamma^i> \frac{Re(\lambda_A^i)}{Re(\lambda^2_\L)}$,  $Re(\lambda_A^i - \gamma^i \lambda_\L^j) < 0$ for all $j \in \mathbb{Z}_2^m$ which implies that
\begin{align}\label{eqn:tinfpmbxi}
\lim_{t \rightarrow \infty} \pmb{x}^i (t)=e^{\lambda_A^i t}\xi_1'\pmb{x}^i(0)1_m
\end{align}
The above is for the single block of $\pmb{x}^i$. Using the same development and noting that $\lambda_A^1$ is the largest eigenvalue of $S$ and that $\lambda^2_{\L}$ is the second smallest eigenvalue of $\L$, the expression is applicable for all $i \in \mathbb{Z}^r$ as  $\lambda_A^1, \cdots, \lambda_A^r$ are unstable eigenvalues. For $i \in \mathbb{Z}_{r+1}^n$, it follows from $Re(\lambda_A^i)<0$ that $\gamma^i=1 > \frac{Re(\lambda_A^i)}{Re(\lambda^2_\L)}$ and $\lim_{t \rightarrow \infty} \pmb{x}^i =0$. This shows $\hat{\pmb{x}}$ reaches consensus.

\noindent $(\Leftarrow)$ Suppose one of the $\gamma^i$ where $i \in \mathbb{Z}^r$ is such that $\gamma^i = \frac{Re(\lambda_A^i)}{Re(\lambda^2_\L)}$, then $Re(\lambda_A^i -\gamma^i \lambda_\L^j)=0$ and it follows from (\ref{eqn:pmbxi}) that either there at least one term such that $e^{(\lambda_A^i -\gamma^i \lambda_\L^j)t}$ is a constant or that there is at least a pair of $j$ such that $e^{(\lambda_A^i -\gamma^i \lambda_\L^j)t}$ is a pair of purely imaginary term. In both of these cases, the last term on the right-hand-side of (\ref{eqn:pmbxi}) does not approach 0 when $t \to \infty$. Hence, consensus is not achieved for this $\pmb{x}^i(t)$. The case of $\gamma^i < \frac{Re(\lambda_A^i)}{Re(\lambda^2_\L)}$ follows the same reasoning and the same conclusion.
 $\square$

%To show that this result also holds for all $\pmb{x}^q, q \in \mathbb{Z}^{n}$, the above block equation can be similarly obtained for $\pmb{x}^q$. Note that $\lambda_A^1$ is the largest eigenvalue of $S$ and that $\lambda^2_{\L}$ is the second smallest eigenvalue of $\L$. If $\gamma^1=\cdots=\gamma^r> \frac{Re(\lambda_A^1)}{Re(\lambda^2_\L)}$, then $Re(\lambda_A^q - \gamma^q \lambda^j_\L) < 0$ for all $q \in \mathbb{Z}^{r}$ and $j \in \mathbb{Z}_2^m$ which implies that $\lim_{t \rightarrow \infty} \pmb{x}^q (t)= e^{\lambda_A^qt}\xi_q'\pmb{x}^q(0)1_m$ for all $q \in \mathbb{Z}^{r}$. For the case of $q > r$, $\gamma^q=1$, $Re(\lambda_A^q) \leq 0$ and $Re(\lambda_A^q - \gamma^q \lambda^j_\L) < 0$  which implies that  $\lim_{t \rightarrow \infty} \pmb{x}^q (t)= 0$. $\square$

The above uses $\Gamma=diag_n\{\gamma^1, \cdots, \gamma^n\}$ with $\gamma^i > \frac{Re(\lambda_A^i)}{Re(\lambda^2_\L)}$ for $i \in \mathbb{Z}^r$ and $\gamma^{r+1}=\cdots=\gamma^n=1$.  A useful choice of $\gamma^i$ that satisfies the consensus condition is to let $\gamma^1=\cdots=\gamma^n > \frac{Re(\lambda_A^1)}{Re(\lambda^2_\L)}$. That this choice of $\gamma$ is sufficient to enforce consensus is easy to see from the proof of Theorem \ref{thm:fixed}. Hence,  the following corollary is stated without proof.
\begin{corollary} \label{cor:1}
Suppose all conditions and variables of Theorem \ref{thm:fixed} hold.  The MAS of (\ref{eqn:dotxiS}) with conditions (\ref{eqn:S}), (\ref{eqn:specAinC}) and (\ref{eqn:Gamma}) reaches consensus if
%(i) $\Gamma=diag_n\{\gamma^1, \cdots, \gamma^n\}$ with $\gamma^1=\cdots=\gamma^r > \frac{Re(\lambda_A^1)}{Re(\lambda^2_\L)}$ and $\gamma^{r+1}=\cdots=\gamma^n=1$. (ii)
$\gamma^1=\cdots=\gamma^n > \frac{Re(\lambda_A^1)}{Re(\lambda^2_\L)}$.
\end{corollary}
\begin{remark} \label{rem:gamma1}
The corollary above has an interesting implication.  If $\Gamma = \gamma^1 I_n$, then $\Phi=Q \gamma^1I_n Q^{-1} = \gamma^1 I_n$. In the original $z$ coordinates, the $i$ agent can be written as
\begin{align}\label{eqn:z2}
  \dot{z}_i &= A z_i + \gamma^1 \sum \a_{ij} (z_j - z_i) = A z_i +  \sum \beta_{ij} (z_j - z_i)
\end{align}
where $\beta_{ij}=\gamma^1 \a_{ij}$. Then a Laplacian matrix $\L_\beta$ exists with $[\L_\beta]_{ij}=\beta_{ij}$ such that $(\ref{eqn:z2})$ reaches consensus. That $\L_\beta$ is a valid Laplacian matrix is easy to verify given that $\L$ is a Laplacian matrix and $\gamma^1 >0$.
\end{remark}

\begin{remark} \label{rem:barz}
The expression of the $z_i$ when $\hat{\pmb{x}}$ reaches consensus is of interest. Since there are $r$ unstable eigenvalues in $A$, it follows from (\ref{eqn:tinfpmbxi}) that only the first $r$ of $\pmb{x}^i$ are non-trivial. Using $z_j =Q x_j$ and (\ref{eqn:tinfpmbxi}), the asymptotic consensus value is
\begin{align}\label{eqn:barzt}
lim_{t \to \infty} z_j(t)&=\sum_{i=1}^r q_i x_j^i(t)= \sum_{i=1}^r q_i  e^{\lambda_A^i t}\mu^i(0)\nonumber\\&=\sum_{i=1}^r q_i  e^{\lambda_A^i t}\xi_1'(I_m \otimes p_i')\pmb{\check{z}}(0), \: j \in \mathbb{Z}_1^m
\end{align}
where $q_i$ is the $i$-th column of $Q$, $\mu^i(0)=\xi_1'\pmb{x}^i(0)$ and $\pmb{x}^i(0)= [p_i'z_1(0) \: p_i' z_2(0) \cdots p_i'z_m(0)]'= (I_m \otimes p_i')\check{\pmb{z}}(0)$ where $\check{\pmb{z}}(0)=[(z_1(0))' \cdots (z_m(0))']'$ and $p_i'$ is the $i^{th}$ row of $Q^{-1}$ following $x_j^i(0)= p_i' z_j(0)$.
\end{remark}

%\begin{remark} \label{rem:barz}
%\rrr need to rewrite to simplify \kkk
%The expression of the $z_i$ when $\hat{\pmb{x}}$ reaches consensus is of interest. Since there are $r$ unstable eigenvalues in $A$, it follows from $(\ref{eqn:tinfpmbxi})$ that $x_j^i(t)= e^{\lambda_A^i t}\mu^i(0)$ for all $i \in \mathbb{Z}_1^n$ and all $j \in \mathbb{Z}_1^m$ with $\mu^i(0)=\xi_1'\pmb{x}^i(0)$. In addition, it follows from $x_j=Q^{-1}z_j$ that $x_j^i(0)= p_i' z_j(0)$ and $\pmb{x}^i(0)= [p_i'z_1(0) \: p_i' z_2(0) \cdots p_i'z_m(0)]'= (I_m \otimes p_i')\pmb{\check{z}}(0)$ where $p_i'$ is the $i^{th}$ row of $Q^{-1}$. However, the $r+1$ to $n$ elements of all $x_j$ approach zero as $t \rightarrow \infty$. Hence, the consensus value of $z_j, \bar{z}_j$, is given by
%\begin{align}\label{eqn:barzt}
%\bar{z}_j(t)=\sum_{i=1}^r q_i x_j^i(t)= \sum_{i=1}^r q_i  e^{\lambda_A^i t}\mu^i(0)=\sum_{i=1}^r q_i  e^{\lambda_A^i t}\xi_1'(I_m \otimes p_i')\pmb{\check{z}}(0), \: j \in \mathbb{Z}_1^m
%\end{align}
%where $q_i$ is the $i$-th column of $Q$.
%\end{remark}

\begin{remark} \label{rem:specialcases}
The above expression of (\ref{eqn:barzt}) is the general expression of the consensus values for system (\ref{eqn:dotzi}) having $r$ unstable eigenvalues. If each $z_i$ is a scalar system (in which case, $z_i=x_i$ and $r=1$) with unstable eigenvalue $\lambda_A^1$ such that $\dot{x}_i=\lambda_A^1 x_i + \gamma \sum_j \alpha_{ij}(x_j-x_i)$, the final consensus value becomes $e^{\lambda_A^1 t}\xi_1'\pmb{x}^i(0)$. In addition, when $\lambda_A^1=0$, the consensus value of $x_i(t)$ becomes the well-known result of $\xi_1'\pmb{x}^i(0)$ for directed graph and $\hat{1}_m'\pmb{x}^i(0)$ for undirected graph.
\end{remark}

\section{Consensus under Switching among Undirected and Connected Graphs} \label{sec:undirected}
This section deals with the MAS of (\ref{eqn:dotxiS}) where $G(t)$ switches among a set of undirected and connected graphs at every $t_k$ as in (A1). Let the collection of these graphs be denoted by
\begin{align}\label{eqn:Iv}
\Omega :=\{ \L^i: i \in \mathbb{Z}^v\}.
\end{align}
Hence, the switching is such that $\L(t) \in \Omega$ for all $t$. A typical representation of the switching process is the use of an indication function
\begin{align}\label{eqn:sigma}
\sigma: \mathbb{R}_0^+ \rightarrow \mathbb{Z}^v
\end{align}
such that $\sigma(t)$ indicates the index $i$ of $\L^i \in \Omega$ at time $t$ and it is continuous on the left. In addition,\\
\textbf{(A5)} Each $\L^i \in \Omega$ corresponds to an undirected graph $G_i$ is connected. \\
\begin{theorem}\label{thm:undirected}
Suppose (A1)-(A5) hold. Consider the system of (\ref{eqn:dotxiS}) with (\ref{eqn:S}), (\ref{eqn:specAinC}) and  (\ref{eqn:Gamma}) holding and arbitrary switching of $G(t)$ such that $\L(t)=\L^{\sigma(t)} \in \Omega$ for all $t$.  Let $\gamma^1=\cdots=\gamma^r$ and $\gamma^{r+1}=\cdots=\gamma^n=1$. Then (\ref{eqn:dotxiS}) reaches consensus exponentially if and only if
\begin{align}
\lambda_A^1 - \gamma^1 \lambda_{\Omega}^2 < 0 \label{eqn:lambdaA1}
\end{align}
where $\lambda_{\Omega}^2:= min \{\lambda^2_{\L^i}: \lambda^2_{\L^i} \textrm{ is the second smallest }$ $\textrm{eigenvalue of } \L^i, i \in \mathbb{Z}^v\}$.
\end{theorem}
\textbf{Proof:}
$(\Rightarrow)$  The proof begins with showing $\pmb{x}^1(t)$ reaches consensus, to be followed by $\pmb{x}^q(t)$ for $q\in \mathbb{Z}_2^n$. Let $\L_k, h_k$ be as defined by (\ref{eqn:Lt}). Since $\L_k$ satisfies (A5) for all $k$, they satisfy (D1)-(D4). The same is true for $\gamma \L_k$ for any $\gamma>0$. Following from (D2) and (D3), $e^{-\gamma \L_k h_k}$ is a symmetric and doubly stochastic matrix with a simple largest eigenvalue of 1 with $\hat{1}_m$ being the corresponding left and right eigenvectors.
Following the same development from (\ref{eqn:dotxiS}) to (\ref{eqn:dotpmbx}),
\begin{align*}
&\dot{\hat{\pmb{x}}}(t_{k+1})= (S \otimes I_m- \Gamma \otimes \L_k) \hat{\pmb{x}}(t_k)\nonumber\\
&= diag_n \{(\lambda^1_AI_m -\gamma^1\L_k), \cdots, (\lambda^n_AI_m-\gamma^n \L_k)\} \hat{\pmb{x}}(t_k)
\end{align*}
and has a solution for $\pmb{x}^1$ as
\begin{align}\label{eqn:ctug3}
&\pmb{x}^1(t_{k+1})=e^{(\lambda^1_A I_m -\gamma^1\L_k) h_k}\pmb{x}^1(t_k)\nonumber=e^{\lambda^1_A I_m h_k -\gamma^1\L_k h_k}\pmb{x}^1(t_k)\nonumber\\
%=&e^{\lambda^1_s h_k}e^{-\L_k h_k}e^{\lambda^1_s h_{k-1}}e^{-\L_{k-1} h_{k-1}}{\pmb{x}}^1(t_{k-1})\nonumber\\
&=e^{\lambda_A^1 h_{k}}e^{\lambda^1_A h_{k-1}} \cdots e^{\lambda^1_A h_{0}} \W(k) \W(k-1)\cdots \W(0){\pmb{x}}^1(t_0)\nonumber\\&=e^{\lambda_A^1 t_{k+1}} \W_0^k\pmb{x}^1(t_0)
\end{align}
where $\W(j)$ and $\W_0^k$ are defined by (\ref{eqn:W}).
%where
%\begin{align}
%\W(j):=e^{- \gamma \L_j h_j}, \textrm{ and } \W_0^k:=\W(k-1)\cdots \W(0). \label{eqn:W}
%\end{align}\kkk
Consider the decomposition of $\W(j)$ by letting
\begin{align*} %\label{eqn:B}
       \D(j):=\W(j)-\hat{1}_m\hat{1}_m', %\quad \textrm{ where } \hat{1}_m :=\frac{1_m}{\sqrt{m}},
\end{align*}
and note that $\D(j)\hat{1}_m=0=\hat{1}_m'\D(j)$ since $\D(j)$ is symmetric. Using these in (\ref{eqn:ctug3}) leads to
\begin{align}
    \pmb{x}^1(t_{k+1})&=e^{\lambda^1_A t_{k+1}}(\D(k)+\hat{1}_m \hat{1}_m') \cdots\nonumber\\& (\D(1)+\hat{1}_m \hat{1}_m')(\D(0)+\hat{1}_m \hat{1}_m')\pmb{x}^1(t_0) \nonumber\\
%      &=e^{\lambda^1_s  t_{k+1}}(\D(k)+\hat{1}_m \hat{1}_m')\cdots(\D(1)\D(0)+\D(1)\hat{1}_m \hat{1}_m'+\hat{1}_m \hat{1}_m'\D(0)+\hat{1}_m \hat{1}_m'\hat{1}_m \hat{1}_m')\pmb{x}^1(t_0)\nonumber\\
      =e^{\lambda^1_A  t_{k+1}}&(\D(k)+\hat{1}_m \hat{1}_m'{m})\cdots(\D(1)\D(0)+\hat{1}_m \hat{1}_m')\pmb{x}^1(t_0) \nonumber\\
      =e^{\lambda^1_A  t_{k+1}}&(\D(k)\D(k-1)\cdots\D(1)\D(0)+\hat{1}_m \hat{1}_m')\pmb{x}^1(t_0) \label{eqn:pmbxupper1}
\end{align}
where the last equality of (\ref{eqn:pmbxupper1}) is the repeated application of the $(\D(k+1)+\hat{1}_m \hat{1}_m')(\D(k)+\hat{1}_m \hat{1}_m')=\D(k+1)\D(k)+ \hat{1}_m \hat{1}_m'$ for all $k$.
Note that each $\L_j$ is connected and symmetric under (A5), hence $spec(\L_j)=\{0, \lambda_{\L_j}^2, \cdots, \lambda_{\L_j}^m\}$ with $0 < \lambda_{\L_j}^2 \leq \cdots \leq \lambda_{\L_j}^m$. It follows from (D4) and (P6) that $spec(\gamma^1\L_j)=\{0, \gamma^1\lambda_{\L_j}^2, \cdots, \gamma^1\lambda_{\L_j}^m\}$ and
$spec(\W(j))=\{1, e^{-\gamma^1\lambda_{\L_j}^2 h_j}, \cdots, e^{-\gamma^1\lambda_{\L_j}^m h_j}\}$.
In addition,
$e^{-\gamma^1\lambda_{\L_j}^q h_j} < 1$ for all $q\in \mathbb{Z}_2^m$ and for all $h_j >0$ since $e^{-\gamma \L_j h_j}$ is a symmetric and doubly stochastic matrix. Similarly, $spec(\D(j))=\{ e^{-\gamma^1\lambda_{\L_j}^2 h_j},$ $e^{-\gamma^1\lambda_{\L_j}^3 h_j}, \cdots, e^{-\gamma^1\lambda_{\L_j}^{m-1} h_j}, 0\}$ and
$\|\D(j) \| = e^{-\gamma^1\lambda_{\L_j}^2 h_j}$. Using this expression of $\|\D(j) \|$, (\ref{eqn:pmbxupper1}) can be further expressed as
\begin{align}
&\pmb{x}^1(t_{k+1})-e^{\lambda^1_A  t_{k+1}}\hat{1}_m \hat{1}_m'\pmb{x}^1(t_0)\nonumber\\&=e^{\lambda^1_A  t_{k+1}}\D(k)\D(k-1)\cdots\D(0)\pmb{x}^1(t_0)\nonumber\\
%     \|\pmb{x}^1(t_{k+1})-e^{\lambda^1_A  t_{k+1}}e_Me_M'\pmb{x}^1(t_0)\|&=e^{\lambda^1_A  t_{k+1}}\D(k)\D(k-1)\cdots\D(0)\pmb{x}^1(t_0)\| \label{eqn:pmbx1}\\
&= (e^{\lambda^1_A h_k}\D(k)) (e^{\lambda^1_A h_{k-1}} \D(k-1)) \cdots (e^{\lambda^1_A h_0}\D(0))\pmb{x}^1(t_0)\label{eqn:elambda1}\\
&\Rightarrow \|\pmb{x}^1(t_{k+1})-e^{\lambda^1_A  t_{k+1}}\hat{1}_m \hat{1}_m'\pmb{x}^1(t_0)\| \nonumber\\&\leq e^{\lambda^1_A h_k}\|\D(k)\| \cdots e^{\lambda^1_A h_0}\|\D(0)\| \|\pmb{x}^1(t_0)\|\nonumber\\
&= e^{ -(\gamma^1\lambda^2_{\L_k}-\lambda^1_A )h_k} \cdots e^{-(\gamma^1\lambda^2_{\L_0}- \lambda^1_A)h_0} \|\pmb{x}^1(t_0)\| \nonumber \\
&\leq e^{-(\gamma^1\lambda^2_\Omega- \lambda^1_A )h_k} \cdots e^{-(\gamma^1\lambda^2_\Omega - \lambda^1_A )h_0}\|\pmb{x}^1(t_0)\|  \label{eqn:leq}\\&= e^{-(\gamma^1\lambda^2_\Omega - \lambda^1_A )t_{k+1}} \|\pmb{x}^1(t_0)\| \nonumber
\end{align}
where the inequality of (\ref{eqn:leq}) follows from that fact that $\lambda_{\Omega}^2= min \{\lambda^2_{\L}: \L \in \Omega\}$. If $\gamma^1\lambda_{\Omega}^2 > \lambda_A^1$, then the right hand side of the above goes to zero exponentially and $\pmb{x}^1(t)$ reaches consensus exponentially towards $e^{\lambda^1_A  t_{k+1}}1_m \alpha(t_0)$ with $\alpha(t_0)= \hat{1}_m'\pmb{x}^1(t_0)$.

The above reasoning is for $\pmb{x}^1(t)$. It can be extended directly for $\pmb{x}^q(t), q\in \mathbb{Z}_2^n$  by noting that $\pmb{x}^q(t_{k+1})=e^{\lambda_A^q t_{k+1}} \W_0^k\pmb{x}^q(t_0)$ from (\ref{eqn:ctug3}) and that $\gamma^1 \lambda_{\Omega}^2 > \lambda_A^1$ implies that $\gamma^1\lambda_{\Omega}^2 > \lambda_A^q$ for $q\in \mathbb{Z}_2^n$. Hence, $\pmb{x}^q(t)$ reaches consensus for $q\in \mathbb{Z}_2^n$ and that $\hat{\pmb{x}}(t)$ reaches consensus exponentially.

$(\Leftarrow)$ The proof is to show that there exists a switching sequence and an initial state $\hat{\pmb{x}}(0)$ such that consensus is not reached when $\gamma^1 \lambda_{\Omega}^2 \leq \lambda_A^1$. Let $i^*= arg min \{\lambda^2_{\L_i}: \L_i \in \Omega, i \in \mathbb{Z}^v\}$ and $\xi_{i^*}$ be the eigenvector of $\D(i^*)$ corresponding to the eigenvalue of $e^{-\gamma^1 \lambda^2_{\L_{i^*}}}$. Since the switching is arbitrary, choose the sequence $\L_k=\L_{i^*}$ for all $k$ and $\pmb{x}^1(t_0)=\xi_{i^*}$, then following the same reasoning given in the proof of the "if" condition up to (\ref{eqn:elambda1}), noting that $\D(k)=\D(k-1)=\cdots=\D(0)=\D(i^*)$, it follows from $\D(k)\xi_{i^*}= e^{-\gamma^1\lambda^2_{\L_{i*}} h_k}\xi_{i^*}$ that
\begin{align*}
    &\pmb{x}^1(t_{k+1})- e^{\lambda^1_At_{k+1}} \hat{1}_m \hat{1}_m'\xi_{i^*}\nonumber\\ &=(e^{\lambda^1_A h_k}\D(k)) (e^{\lambda^1_A h_{k-1}} \D(k-1)) \cdots (e^{\lambda^1_A h_0}\D(0)) \xi_{i^*}\nonumber\\&=e^{-(\gamma^1 \lambda^2_{\L_{i^*}}-\lambda^1_A)t_{k+1}}\xi_{i^*}
\end{align*}
When $\lambda_A^1= \gamma^1 \lambda^2_{\L_{i^*}}$, the right hand side is $\xi_{i^*}$. Since $\xi_{i^*}$ is not a scalar multiple of $1_m$ (eigenvector of $e^{-\lambda^2_{\L_{i^*}}h_j}$ is orthogonal to $1_m$), $\pmb{x}^1(t)$ does not reach consensus. $\square$

\section{Consensus under Switching among Directed and Connected Graphs} \label{sec:directed}
The result of the above is now extended to the case where the graphs are directed and connected. For this purpose, the set of $\Omega=\{ \L^1, \cdots, \L^v\}$
is such that \\
\textbf{(A6)} Each $\L^i \in \Omega$ corresponds to a strongly connected directed graph $G_i$.\\
\begin{theorem}\label{thm:directed}
Suppose (A1)-(A4) and (A6) hold. Consider the system of (\ref{eqn:dotxiS}) with (\ref{eqn:S}), (\ref{eqn:specAinC}) and (\ref{eqn:Gamma}) holding. Let $\gamma^1=\cdots=\gamma^r$ and $\gamma^{r+1}=\cdots=\gamma^n=1$. Then, (\ref{eqn:dotxiS}) reaches consensus exponentially when $\bar{\gamma}$ is sufficiently large.
\end{theorem}
\textbf{Proof:} The expression of $\pmb{x}^1(t_{k+1})$ is the same as those in the proof of Theorem \ref{thm:undirected} until equation (\ref{eqn:ctug3}).
Let $\B(j):=\W(j)- \hat{1}_m\xi'_j$ where $\xi_j'$ with $\xi_j' \hat{1}_m=1$ is the left eigenvector of $\W(j)$ corresponding to eigenvalue 1. Then (\ref{eqn:ctug3}) can be written as
\begin{align}
 \pmb{x}^1(t_{k+1})=e^{\lambda_A^1 t_{k+1}}&(\B(k)+\hat{1}_m\xi'_k)\cdots\nonumber\\ &(\B(1)+\hat{1}_m\xi'_1)(\B(0)+\hat{1}_m\xi'_0)\pmb{x}^1(t_0)\nonumber\\
 =e^{\lambda_A^1t_{k+1}}&(\B(k)+\hat{1}_m\xi'_k)\cdots\nonumber\\ &(\B(1)\B(0)+\hat{1}_m(\xi'_1\B(0)+\xi'_0))\pmb{x}^1(t_0) \nonumber
\end{align}
since $\W(1)\hat{1}_m= \hat{1}_m$ and $(\B(1)+\hat{1}_m\xi'_1)(\B(0)+\hat{1}_m\xi'_0)
=\B(1)\B(0)+\hat{1}_m\xi'_1\B(0)+(\W(1)-\hat{1}_m\xi'_1)\hat{1}_m\xi'_0+\hat{1}_m\xi'_1\hat{1}_m\xi'_0
=\B(1)\B(0)+\hat{1}_m\xi'_1\B(0)+\hat{1}_m\xi'_0=\B(1)\B(0)+\hat{1}_m(\xi'_1\B(0)+\xi'_0)$.
%\begin{align*}
%(\B(1)+\hat{1}_m\xi'_1)(\B(0)+&\hat{1}_m\xi'_0)
%=\B(1)\B(0)+\hat{1}_m\xi'_1\B(0)+\nonumber\\
%&(\W(1)-\hat{1}_m\xi'_1)\hat{1}_m\xi'_0+\hat{1}_m\xi'_1\hat{1}_m\xi'_0\\
%=\B(1)\B(0)+\hat{1}_m&\xi'_1\B(0)+\hat{1}_m\xi'_0\nonumber\\ =\B(1)\B(0)+\hat{1}_m&(\xi'_1\B(0)+\xi'_0)
%\end{align*}
Repeating this process for $\B(0)$ to $\B(k)$ yields
\begin{align}
&\pmb{x}^1(t_{k+1})-e^{\lambda_A^1 t_{k+1}}\hat{1}_m\bar{\xi}'_{k}\pmb{x}^1(0)\nonumber\\&=e^{\lambda_A^1 t_{k+1}}(\B(k)\B(k-1)\cdots\B(0))\pmb{x}^1(t_0)\nonumber\\
&=(e^{\lambda_A^1 h_k}\B(k))(e^{\lambda_A^1 h_{k-1}}\B(k-1)\cdots (e^{\lambda_A^1 h_0}\B(0))\pmb{x}^1(t_0)\nonumber
\end{align}
where $\bar{\xi}'_{k}={\xi}'_{0}+{\xi}'_{1}\B(0)+{\xi}'_{2}\B(1)\B(0)+\cdots+{\xi}'_{k}\B(k-1)\B(k-2)\cdots\B(0)$. The above also implies that
\begin{align}
\|\pmb{x}^1(t_{k+1})&-e^{\lambda_A^1 t_{k+1}}\hat{1}_m\bar{\xi}'_{k}\pmb{x}^1(0)\|\nonumber\\&\leq\|e^{\lambda_A^1 h_k}\B(k)\|\cdots\|e^{\lambda_A^1 h_0}\B(0)\|\|\pmb{x}^1(t_0)\|\label{eqn:lambdas1hkBk}
\end{align}
Consider the eigen-decomposition of $\W(j)$ in the form of
\begin{align}
\W(j)=\sum_{i=1}^m q_i p_i' e^{-\gamma^1 \lambda_{\L_j}^i h_j} \label{eqn:barWqp}
\end{align}
where $q_i, p_i$ are the right and left eigenvectors of $\W(j)$ corresponding to eigenvalue $e^{-\gamma^1 \lambda_{\L_j}^i h_j}$. It follows from (\ref{eqn:barWqp}) that $\B(j) = \sum_{i=2}^{m} q_ip_i e^{- \gamma^1\lambda_{\L_j}^i h_j}$. Hence,
\begin{align*}
\|e^{\lambda_A^1h_j}\B(j)\|& \leq \sum_{i=2}^{m} \| q_ip_i\|  e^{- (\gamma^1\lambda_{\L_j}^i  - \lambda_A^1)h_j}.\\
& \le \sum_{i=2}^{m} \| q_ip_i\| e^{- (\gamma^1\lambda_{\L_j}^i  - \lambda_A^1)\underline{h}}.
% & \leq \sum_{i=2}^{m} \| q_ip_i \| \|e^{-(\gamma^1\lambda_{\L_j}^i-\lambda_A^1)h_j}\|
\end{align*}  \kkk
where $\underline{h}$ is that given in (A4). Clearly, when $\gamma^1\to\infty$, $\| e^{\lambda_A^1h_j}\B(j)\| \to 0$. This holds also for
all $\|e^{\lambda_A^1h_i}\B(i)\|, i=0,\cdots, k$. Hence, the RHS of (\ref{eqn:lambdas1hkBk}) approaches $0$ and $\pmb{x}^1(t_{k+1}) \to e^{\lambda_A^1t_{k+1}}\hat{1}_m\bar{\xi}_k'\pmb{x}^1(0)$.
Repeating the above to $\pmb{x}^q(t_{k+1})$ for $q=2, \cdots, n$ completes the proof. $\square$

%It is noteworthy that the results of Theorem \ref{thm:undirected} are both necessary and sufficient when $\Gamma$ is a diagonal matrix. In contrast, results of Theorem \ref{thm:directed} is only sufficient.

\begin{remark}
It is of interest to note that both Theorem \ref{thm:undirected} and \ref{thm:directed} are achieved with $\underline{h}>0$ of (A4). This is a mild requirement on the dwell time in each mode.
%A similar conclusion is reached by \cite{ART:VZ17} using Lyapunov stability theory for the single input case.
\end{remark}
\kkk

\section{Numerical Example}\label{sec:num}
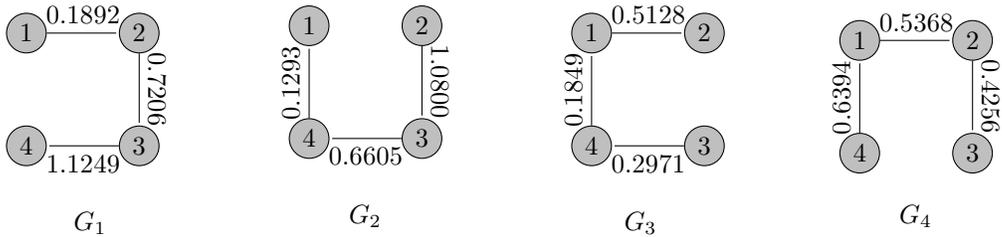
\begin{figure}[h]
    \centering
    \begin{minipage}{0.2\textwidth}
    \centering
        \begin{tikzpicture}[-,shorten >=1pt,node distance=1.5cm,on grid,auto]
            \tikzstyle{every state}=[circle,fill=black!25,minimum size=15pt,inner sep=0pt]
            \node[state](1){1};
            \node[state](2)[right of=1]{2};
            \node[state](3)[below of=2]{3};
            \node[state](4)[left of=3]{4};
            \path (1) edge node {$0.1892$}(2)
            (2) edge node [sloped, anchor=center, above, text width=1.0cm] {$0.7206$}(3)
            (3) edge node {$1.1249$}(4);
        \end{tikzpicture}\\\textbf{$G_1$}
    \end{minipage}
    \begin{minipage}{0.2\textwidth}
    \centering
            \begin{tikzpicture}[-,shorten >=1pt,node distance=1.5cm,on grid,auto]
            \tikzstyle{every state}=[circle,fill=black!25,minimum size=15pt,inner sep=0pt]
            \node[state](1){1};
            \node[state](2)[right of=1]{2};
            \node[state](3)[below of=2]{3};
            \node[state](4)[left of=3]{4};
            \path (4) edge node [sloped, anchor=center, above, text width=1cm] {$0.1293$}(1)
            (2) edge node [sloped, anchor=center, above, text width=1.0cm] {$1.0800$}(3)
            (3) edge node {$0.6605$}(4);

        \end{tikzpicture}\\\textbf{$G_2$}
    \end{minipage}
    \begin{minipage}{0.2\textwidth}
    \centering
        \begin{tikzpicture}[-,shorten >=1pt,node distance=1.5cm,on grid,auto]
            \tikzstyle{every state}=[circle,fill=black!25,minimum size=15pt,inner sep=0pt]
            \node[state](1){1};
            \node[state](2)[right of=1]{2};
            \node[state](3)[below of=2]{3};
            \node[state](4)[left of=3]{4};
            \path (4) edge node [sloped, anchor=center, above, text width=1.0cm] {$0.1849$}(1)
            (1) edge node {$0.5128$}(2)
            (3) edge node {$0.2971$}(4);
        \end{tikzpicture}\\\textbf{$G_3$}
    \end{minipage}
    \begin{minipage}{0.2\textwidth}
    \centering
\begin{tikzpicture}[-,shorten >=1pt,node distance=1.5cm,on grid,auto]
            \tikzstyle{every state}=[circle,fill=black!25,minimum size=15pt,inner sep=0pt]
            \node[state](1){1};
            \node[state](2)[right of=1]{2};
            \node[state](3)[below of=2]{3};
            \node[state](4)[left of=3]{4};
            \path (4) edge node [sloped, anchor=center, above, text width=1.0cm] {$0.6394$}(1)
            (1) edge node {$0.5368$}(2)
            (2) edge node [sloped, anchor=center, above, text width=1.0cm] {$0.4256$}(3);
        \end{tikzpicture}\\\textbf{$G_4$}
    \end{minipage}
    \caption{Interaction among agents}
    \label{fig:G_un_directed}
 \end{figure}

An example for the switching among undirected and connected networks is shown. It is for the case of $m=4, n=2$ with each agent of (\ref{eqn:dotziopenloop}) having a dynamic compensator of (\ref{eqn:doteta1}) -(\ref{eqn:doteta2}). The matrices are
$$A=\left[\begin{array}{cc}
     -1.5 & 2 \\
     -1.28& 1.7
   \end{array}\right],\quad
B=[1\:2]',\quad K=[0.1333\;-1.9167]$$ and the transformation matrix $$Q=\left[\begin{array}{cc}-0.2&-0.5\\-.16&-0.5\end{array}\right].$$
Initial conditions are $w_1(0)=[6 \; -8]',\: w_2(0)=[-12\;6]',\: w_3(0)=[-17\;22]'$, $w_4(0)=[18\;-3]'$ and $\eta_i(0)=0.5 w_i(0)$ for $i \in \mathbb{Z}^4$.
%$\eta_1(0)=0.5w_1(0),\:\eta_2(0)=0.5w_2(0),\:\eta_3(0)=0.5w_3(0)$, $\eta_4(0)=0.5w_4(0)$.
Note that $A$ is unstable with $J=Q^{-1}AQ=0.1 I_2 + \N^2$. The communication network switches arbitrarily among four undirected and connected graphs with topologies and weights given in Figure \ref{fig:G_un_directed}. The dwell times, $h_k$, are sampled from a uniformly distribution within the interval $[0.5,1]$ while $\sigma(t)$ is drawn from a uniform distribution of the four modes.

\begin{figure}
\centering
\begin{subfigure}{0.4\textwidth}
\includegraphics[width=1\textwidth]{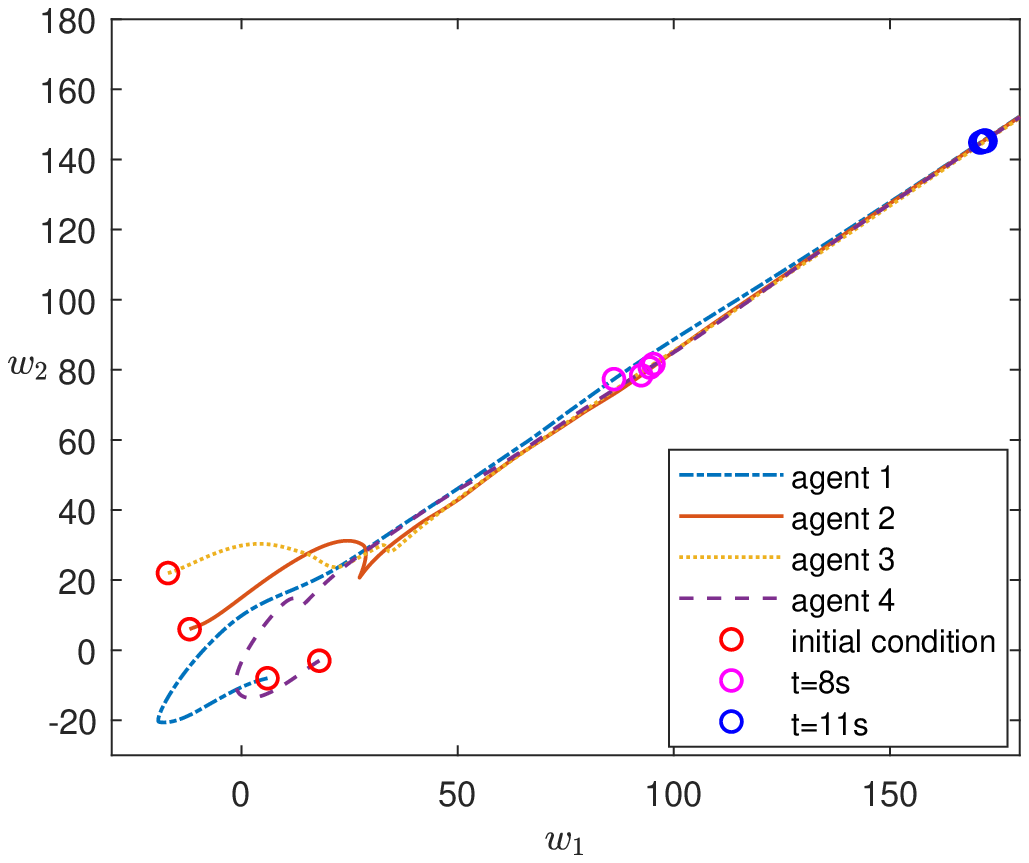}
\caption{}
\label{fig:uc_z1_z2}
\end{subfigure}
\begin{subfigure}{0.4\textwidth}
\includegraphics[width=1\textwidth]{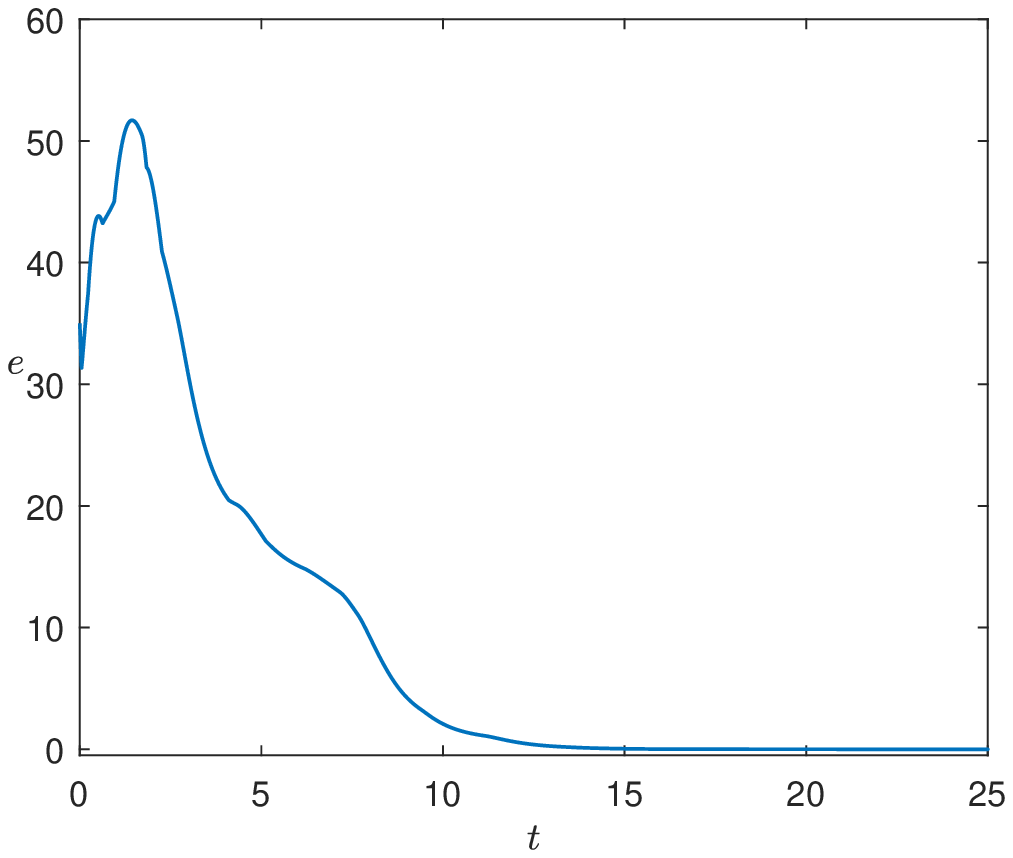}
\caption{}
\label{fig:uc_error}
\end{subfigure}
\caption{ Plots of (\subref{fig:uc_z1_z2}) $w_1(t)$ versus $w_2(t)$, (\subref{fig:uc_error}) absolute error values between agents, $ e(t):=max\{\|w_i(t)-w_j(t)\|_\infty,\:(i,j)\in\E\}$, versus $t$.}
\end{figure}

\begin{figure}
\includegraphics[width=0.4\textwidth]{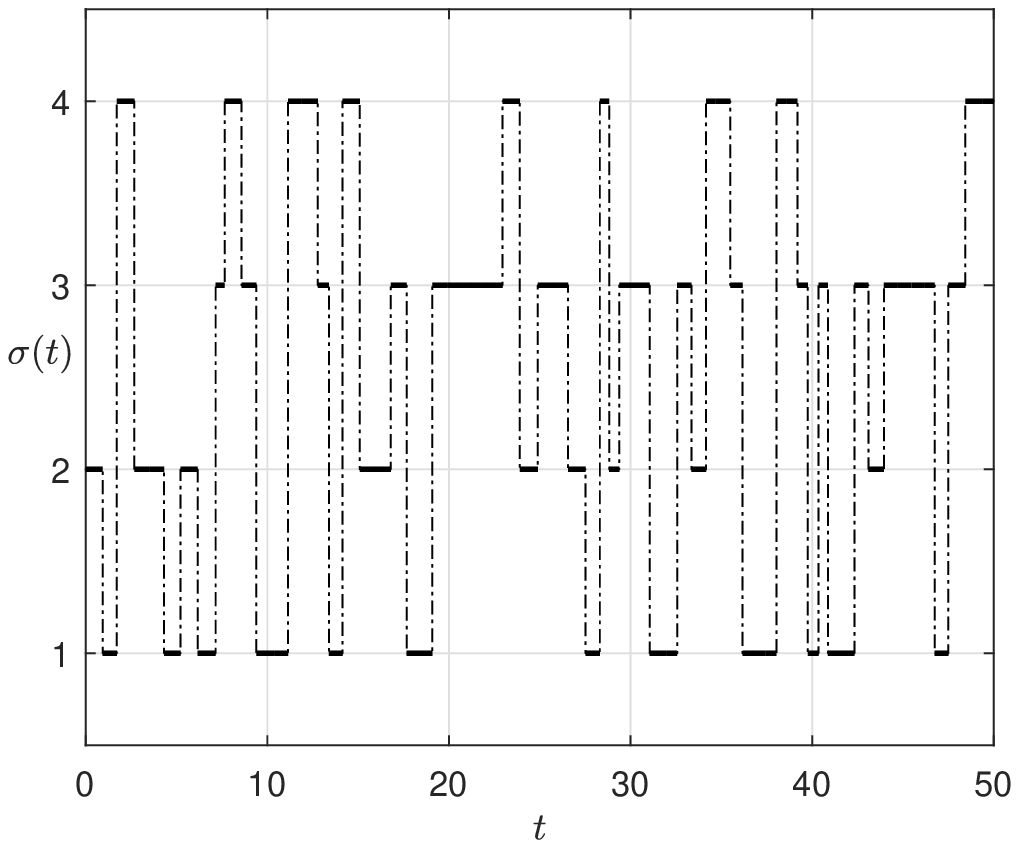}
\caption{ Plot of $\sigma(t)$ versus $t$.}
\label{fig:uc_sigma}
\end{figure}

The matrix $\Phi=Q\Gamma Q^{-1}=\left[\begin{array}{cc}6.5&-5\\4&-2.5\end{array}\right]$ is obtained from $\Gamma=diag_2\{2.5,\:1.5\}$
which satisfies condition (\ref{eqn:lambdaA1}) of Theorem \ref{thm:undirected}. Figure \ref{fig:uc_z1_z2} shows the state-space trajectories of the dynamics $w_i$ of \eqref{eqn:dotziopenloop}-\eqref{eqn:doteta2} for $i\in\mathbb{Z}^4$ while Figure \ref{fig:uc_error} shows the maximum error $e(t)=max\{\|\o_i(t)-\o_j(t)\|_\infty | (i,j) \in \E \}$ against $t$. As shown in Figure \ref{fig:uc_z1_z2}, $w_i(t)$ is different at $t=0$ for the four agents but reach consensus after $t=16s$. Figure \ref{fig:uc_sigma} shows the values of $\sigma(t)$.
\kkk
\kkk

\section{Conclusions}\label{sec:con}
This work considers the consensus problem of multi-agent system with identical agents under switching networks. Each agent is described by a general linear dynamical system with possibly repeated unstable eigenvalues. The proposed controller structure is based on a diagonal gain matrix used to achieve consensus in a leaderless setting and does not rely on Lyapunov analysis to establish consensus. Three cases of the communication graphs are considered: (i) fixed; (ii) switching among connected and undirected; and (iii) switching among connected and directed.  The focus is on switching network and the proposed approach achieves consensus uses properties of stochastic matrices and eigendecomposition. Consequently, it yields some interesting results - consensus is achieved so long as the switching is not instantaneous and that the gain values are sufficiently large.

\section{Appendix} \label{sec:Appendix}
\textbf{Proof of Lemma \ref{lem:dotxi}:} Using (\ref{eqn:dothatpmbx}), the system of (\ref{eqn:dotxi3}) can be written as  $\dot{\hat{\pmb{x}}}(t)=(\lambda I_n \otimes I_m -  I_n \otimes \gamma\L(t))\hat{\pmb{x}}(t)$. Since $\lambda I_n \otimes I_m$ and $I_n \otimes \gamma\L(t)$ commute,
\begin{align}
  \hat{\pmb{x}}(t_{k+1})&=e^{\lambda I_n \otimes I_m h_k}e^{ I_n \otimes (-\gamma\L_k h_k)}\hat{\pmb{x}}(t_k) \quad\nonumber\\& (\textrm{using (P3) and notations of (\ref{eqn:Lt})) }\nonumber\\
  &=(e^{\lambda I_n h_k} \otimes I_m)(I_n \otimes e^{-\gamma \L_k h_k})\hat{\pmb{x}}(t_k) \quad\nonumber\\&  (\textrm{using (P4) and (P5))} \nonumber\\
  &=(e^{\lambda I_n h_k} \otimes e^{- \gamma \L_k h_k})\hat{\pmb{x}}(t_k)  \quad (\textrm{using (P1))}\nonumber\\
&=(e^{\lambda h_k} I_n \otimes e^{- \gamma \L_k h_k})\hat{\pmb{x}}(t_k)\nonumber\\&=e^{\lambda h_k} diag_n\{e^{- \gamma \L_k h_k}, \cdots, e^{- \gamma \L_k h_k}\} \hat{\pmb{x}}(t_k)\label{eqn:hatpmbx3}
\end{align}
Clearly, the above is a block-diagonal system and can be expressed as collection of $n$ sets of $m$ equations. For each $j \in \mathbb{Z}^n$ with
$\pmb{x}^j=[x^j_1 \cdots x^j_m]'$, each of the $n$ sets can be expressed as
\begin{align}
  {\pmb{x}}^j(t_{k+1})&=e^{\lambda h_k}e^{-\gamma \L_k h_k}{\pmb{x}}^j(t_k)\nonumber\\&=e^{\lambda h_k}e^{-\gamma \L_k h_k}e^{\lambda h_{k-1}}e^{-\gamma \L_{k-1} h_{k-1}}{\pmb{x}}^j(t_{k-1})\nonumber\\
%  & \vdots\nonumber\\
%  &=e^{\lambda (h_k+h_{k-1}+ \cdots +h_0)}e^{-\L_k h_k}e^{-\L_{k-1} h_{k-1}} \cdots e^{-\L_0 h_0}{\pmb{x}}^j(t_0)\nonumber\\
  &:=e^{\lambda t_{k+1}} \W(k)\W(k-1)\cdots \W(0){\pmb{x}}^j(t_0)\nonumber\\&:=e^{\lambda t_{k+1}} \W_0^k{\pmb{x}}^j(t_0)\label{eqn:pmbxj}
\end{align}
where
\begin{align}
\W(j):=e^{- \gamma \L_j h_j}, \textrm{ and } \W_0^k:=\W(k-1)\cdots \W(0). \label{eqn:W}
\end{align}
If (\ref{eqn:dotxi3}) reaches consensus exponentially for all $\pmb{x}^j(0)$, this means, from (\ref{eqn:pmbxj}), that
\begin{align}
\W_0^k=1_m \psi' + \delta(t_{k+1}) \textrm{ with } \|\delta(t)\| < \delta_0 e^{-\mu t} \label{eqn:W0K}
\end{align}
for some $\psi \in \mathbb{R}^m, \mu, \delta_0>0$ and $lim_{k \rightarrow \infty} \W_0^k=1_m \psi'$.

Now, rewriting (\ref{eqn:dotxit}) using (\ref{eqn:dothatpmbx}) and  (\ref{eqn:dothatpmbx3}) yields
  \begin{align}
  \dot{\hat{\pmb{x}}}(t)=(J_n \otimes I_m - \gamma I_n \otimes \L(t))\hat{\pmb{x}}(t) \label{eqn:dothatpmbx2}
  \end{align}
Note that $(J_n \otimes I_m)( \gamma I_n \otimes \L(t))=  \gamma  J_n \otimes \L(t) = ( \gamma I_n J_n) \otimes ( \L(t) I_m)=( \gamma I_n \otimes \L(t))(J_n \otimes I_m)$ using (P2).
This commutative property means that (P3) can be used in (\ref{eqn:dothatpmbx2}) and (\ref{eqn:hatpmbx}) giving
\begin{align}
  &\hat{\pmb{x}}(t_{k+1})=e^{J_n \otimes I_mh_k}e^{- \gamma I_n \otimes \L_k h_k}\hat{\pmb{x}}(t_k)\nonumber\\
  &=(e^{J_n h_k} \otimes I_m)(I_n \otimes e^{- \gamma \L_k h_k})\hat{\pmb{x}}(t_k)\nonumber\\
  &=(e^{J_n h_k} \otimes e^{- \gamma \L_k h_k})\hat{\pmb{x}}(t_k)\nonumber\\
  &= e^{\lambda h_k} e^{\N^n h_k} \otimes e^{- \gamma \L_k h_k}\hat{\pmb{x}}(t_k) \nonumber\\
  & =e^{\lambda h_k}( e^{\N^n h_k} \otimes e^{- \gamma \L_k h_k})\nonumber\\
&\quad\quad e^{\lambda h_{k-1}} ( e^{\N^n h_{k-1}} \otimes e^{- \gamma \L_{k-1} h_{k-1}})\hat{\pmb{x}}(t_{k-1}) \nonumber\\
  &= e^{\lambda (h_k+h_{k-1})}(e^{\N^n h_k} \otimes e^{- \gamma \L_k h_k})\nonumber\\
&\quad\quad(e^{\N^n h_{k-1}} \otimes e^{- \gamma \L_{k-1} h_{k-1}})\hat{\pmb{x}}(t_{k-1})\nonumber\\
  &= e^{\lambda (h_k+h_{k-1})}( e^{\N^n (h_k+h_{k-1})} \otimes e^{- \gamma \L_k h_k}e^{- \gamma \L_{k-1} h_{k-1}})\nonumber\\
&\quad\quad\hat{\pmb{x}}(t_{k-1}) \textrm{ (using (P2))}\nonumber
\end{align}
Using the same notation for  $\W(j)$ and $\W_0^k$ of (\ref{eqn:W}), repeated application of the above yields
\begin{align}
&\hat{\pmb{x}}(t_{k+1})=e^{\lambda t_{k+1}}e^{\N^n (t_{k+1})}  \otimes \W(k) \W(k-1)\cdots \W(0) \hat{\pmb{x}}(t_0) \nonumber\\&=e^{\lambda t_{k+1}}\begin{bmatrix}
                               \W_0^k & t_{k+1} \W_0^k & \cdots & \frac{ t_{k+1}^{n-1}}{(n-1)!}\W_0^k \\
                               0 & \W_0^k & \cdots & \cdots \\
                               \vdots &  & \ddots &  \\
                               0 & \cdots & \cdots & \W_0^k
                             \end{bmatrix}\hat{\pmb{x}}(t_0) \nonumber \\%\label{eqn:pmbtkplus12} \\
                             &=e^{\lambda t_{k+1}}\begin{bmatrix}
                               I_m & t_{k+1} I_m & \cdots & \frac{t_{k+1}^{n-1}}{(n-1)!}I_m \\
                               0 & I_m & \cdots & \cdots \\
                               \vdots &  & \ddots &  \\
                               0 & \cdots & \cdots & I_m
                             \end{bmatrix} \left(
                                             \begin{array}{c}
                                               \W_0^k \pmb{x}^1(t_0) \\
                                               \vdots \\
                                               \W_0^k \pmb{x}^{n-1}(t_0)\\
                                               \W_0^k \pmb{x}^n(t_0) \\
                                             \end{array}
                                           \right) \label{eqn:hatpmbxtkplus1}
\end{align}
Since $\W_0^k$ approaches consensus exponentially from (\ref{eqn:W0K}) and $\lim_{t \rightarrow \infty} t^r e^{-\mu t}=0$ for any $\mu >0$, it is easy to see from (\ref{eqn:hatpmbxtkplus1}) that $\pmb{x}^j(t)$ reaches consensus for all $j \in \mathbb{Z}^n$. Specifically, $lim_{k \rightarrow \infty} \pmb{x}^j(t_{k+1}) = \lim_{k \rightarrow \infty}  \W_0^k \pmb{x}^j(t_0) + t_{k+1}\W_0^k \pmb{x}^{j+1}(t_0) + \cdots + \frac{t_{k+1}^{n-j}}{(n-j)!} \W_0^k \pmb{x}^n(t_0)$= $1_m (\alpha_j + t_{k+1}\alpha_{j+1} + \cdots + \frac{t_{k+1}^{n-j}}{(n-j)!}\alpha_n)$ since $\W_0^k$ is given by (\ref{eqn:W0K}) and   $\alpha_j= \psi'\pmb{x}^j(t_0)$. Repeating this for all $j \in \mathbb{Z}^n$ shows that $\hat{\pmb{x}}$ reaches consensus $\square$.

\small
\bibliographystyle{unsrtnat}
\bibliography{Reference}

\begin{thebibliography}{13}
\providecommand{\natexlab}[1]{#1}
\providecommand{\url}[1]{\texttt{#1}}
\expandafter\ifx\csname urlstyle\endcsname\relax
  \providecommand{\doi}[1]{doi: #1}\else
  \providecommand{\doi}{doi: \begingroup \urlstyle{rm}\Url}\fi

\bibitem[Ren and Beard(2005)]{ART:RB05}
W.~Ren and R.~W. Beard.
\newblock Consensus seeking in multiagent systems under dynamically changing
  interaction topologies.
\newblock \emph{IEEE Transactions on Automatic Control}, 50\penalty0
  (5):\penalty0 655--661, 2005.

\bibitem[Moreau(2005)]{ART:M05}
L.~Moreau.
\newblock Stability of multiagent systems with time-dependent communication
  links.
\newblock \emph{IEEE Trans. Automatic Control}, 50\penalty0 (2):\penalty0
  169--182, 2005.

\bibitem[Moreau(2004)]{INP:M04}
L.~Moreau.
\newblock Stability of continuous-time distributed consensus algorithms.
\newblock In \emph{2004 43rd IEEE Conference on Decision and Control (CDC)},
  volume~4, pages 3998--4003, 2004.

\bibitem[Ren(2008)]{ART:R08}
W.~Ren.
\newblock On consensus algorithms for double-integrator dynamics.
\newblock \emph{IEEE Trans. Automatic Control}, 53\penalty0 (6):\penalty0
  1503--1509, 2008.

\bibitem[Scardovi and Sepulchre(2009)]{ART:SS09}
L.~Scardovi and R.~Sepulchre.
\newblock Synchronization in networks of identical linear systems.
\newblock \emph{Automatica}, 45\penalty0 (11):\penalty0 2557--2562, 2009.

\bibitem[Tuna(2009)]{ART:T09}
S.~Emre Tuna.
\newblock Conditions for synchronizability in arrays of coupled linear systems.
\newblock \emph{IEEE Transactions on Automatic Control}, 54\penalty0
  (10):\penalty0 2416--2420, 2009.

\bibitem[Han et~al.(2013)Han, Wang, Lin, and Zheng]{ART:HWJ13}
Z.~Han, L.~Wang, Z.~Lin, and R.~Zheng.
\newblock Distributed output regulation of leader–follower multi-agent
  systems.
\newblock \emph{International Journal of Robust and Nonlinear Control},
  23\penalty0 (1):\penalty0 48--66, 2013.

\bibitem[Valcher and Zorzan(2017)]{ART:VZ17}
Maria~Elena Valcher and Irene Zorzan.
\newblock On the consensus of homogeneous multi-agent systems with arbitrarily
  switching topology.
\newblock \emph{Automatica}, 84:\penalty0 79--85, 2017.
\newblock ISSN 0005-1098.

\bibitem[Zhang et~al.(2011)Zhang, Lewis, and Das]{ART:ZLD11}
H.~Zhang, F.~L. Lewis, and A.~Das.
\newblock Optimal design for synchronization of cooperative systems: State
  feedback, observer and output feedback.
\newblock \emph{IEEE Transactions on Automatic Control}, 56\penalty0
  (8):\penalty0 1948--1952, 2011.

\bibitem[Liberzon and Morse(1999)]{ART:LM99}
D.~Liberzon and A.S. Morse.
\newblock Basic problems in stability and design of switched systems.
\newblock \emph{IEEE Control Systems Magazine}, 19\penalty0 (5):\penalty0
  59--70, 1999.

\bibitem[Lin and Antsklis(2009)]{ART:LA09a}
H.~Lin and P.~J. Antsklis.
\newblock Stability and stablizability of switched linear systems: A survey of
  recent results.
\newblock \emph{IEEE Trans. Automatic Control}, 54\penalty0 (3):\penalty0
  308--322, 2009.

\bibitem[Ong and Canyakamz(2022)]{INP:OC22}
C.~J. Ong and I.~Canyakamz.
\newblock Consensus of homogeneous agents with general linear dynamics.
\newblock In \emph{Proceedings of 61th Conference on Decision and Control},
  2022.

\bibitem[Horn and Johnson(1991)]{BOO:HJ91}
R.~A. Horn and C.~R. Johnson.
\newblock \emph{Topics in Matrix Analysis}.
\newblock Cambridge University Press, 1991.

\end{thebibliography}

%% Authors are advised to submit their bibtex database files. They are
%% requested to list a bibtex style file in the manuscript if they do
%% not want to use model2-names.bst.

%% References without bibTeX database:

% \begin{thebibliography}{00}

%% \bibitem must have one of the following forms:
%%   \bibitem[Jones et al.(1990)]{key}...
%%   \bibitem[Jones et al.(1990)Jones, Baker, and Williams]{key}...
%%   \bibitem[Jones et al., 1990]{key}...
%%   \bibitem[\protect\citeauthoryear{Jones, Baker, and Williams}{Jones
%%       et al.}{1990}]{key}...
%%   \bibitem[\protect\citeauthoryear{Jones et al.}{1990}]{key}...
%%   \bibitem[\protect\astroncite{Jones et al.}{1990}]{key}...
%%   \bibitem[\protect\citename{Jones et al., }1990]{key}...
%%   \harvarditem[Jones et al.]{Jones, Baker, and Williams}{1990}{key}...
%%

% \bibitem[ ()]{}

% \end{thebibliography}

\end{document}